\documentclass[prb,twocolumn,superscriptaddress,longbibliography]{revtex4-2}
\usepackage{amsmath,amssymb,latexsym,epsfig,graphics,epsf,subcaption}
\usepackage{xcolor,hyperref}
\hypersetup{
	colorlinks,
	linkcolor={blue!50!black},
	citecolor={blue!50!black},
	urlcolor={blue!80!black}
}
\usepackage{float}
\newcommand{\fig}[1]{Fig.~\ref{#1}}
\newcommand{\Fig}[1]{Figure~\ref{#1}}
\newcommand{\eq}[1]{Eq.~(\ref{#1})}
\newcommand{\Eq}[1]{Equation~(\ref{#1})}

\newcommand{\br}{\bold r}

\newcommand{\be}{\begin{equation}}
\newcommand{\ee}{\end{equation}}

\newcommand{\am}{{\alpha}_{\rm min}}

\newcommand{\Dp}{D_{\rm par}}
\newcommand{\Dh}{D_{\rm heat}}
\newcommand{\Dm}{D_{\rm mom}}
\newcommand{\Dc}{D_{\rm coh}}
\newcommand{\ls}{l_{\rm s}}
\newcommand{\stf}{\textit{solid-that-flows}\,}
\newcommand{\Stf}{\textit{Solid-that-Flows}\,}
\newcommand{\str}{\sigma_{\mu \nu}}
\newcommand{\ta}{\tau_{\alpha}}

\begin{document}
\title{\Stf Picture of Glass-Forming Liquids}
\author{Jeppe C. Dyre}
\affiliation{``Glass and Time'', IMFUFA, Dept. of Sciences, Roskilde University, P. O. Box 260, DK-4000 Roskilde, Denmark}

\date{\today}

\begin{abstract}
This perspective article reviews arguments that glass-forming liquids are different from those of standard liquid-state theory, which typically have a viscosity in the mPa$\cdot$s range and relaxation times on the order of picoseconds. These numbers grow dramatically and become $10^{12}-10^{15}$ times larger for liquids cooled toward the glass transition. This translates into a qualitative difference, and below the ``solidity length'' which is roughly one micron at the glass transition, a glass-forming liquid behaves much like a solid. Recent numerical evidence for the solidity of ultraviscous liquids is reviewed, and experimental consequences are discussed in relation to dynamic heterogeneity, frequency-dependent linear-response functions, and the temperature dependence of the average relaxation time.
\end{abstract}

\maketitle

Liquids flow and solids do not, according to the conventional wisdom. In reality, any solid does flow when subjected to an external force \cite{leb68,sau10,saw16,ton20,fur21,bag22,tu23} while, on the other hand, an extremely viscous liquid only flows very slowly. Should one think of the latter as an ordinary liquid like water or a molten metal, merely with a much higher viscosity, or more as a solid that flows \cite{dyr06,dyr07a}? This question is important for liquids approaching the glass transition where the viscosity is $10^{12}-10^{15}$ times larger than that of ordinary liquids.

A glass is usually made by supercooling a liquid fast enough to avoid crystallization. It is a solid that has inherited the liquid's disorder and macroscopic isotropy. While some substances like pure metals require extremely high cooling rates to form glasses, others, e.g., many organic liquids, are easily supercooled and in fact often difficult to crystallize. Because all substances can form glasses, glass may be regarded as the fourth state of conventional matter \cite{dyr06}.

With only few exceptions like the silicates, a liquid's viscosity $\eta$ at the melting temperature $T_m$ is within one or two orders of magnitude of that of water, $\eta\sim 10^{-3}$ Pa$\cdot$s. Upon supercooling the viscosity increases dramatically, and for typical cooling rates of order K/min one finds $\eta\sim 10^{12}$ Pa$\cdot$s at the glass transition temperature $T_g$ (brief introductions to the glass transition are given in Refs.  \onlinecite{joh74,ang95,deb01,dyr06,alb22}, more comprehensive reviews in Refs. \onlinecite{har76,bra85,varshneya,edi96,ang00,ber11,hun12,wan12,mck17,tan19,mauro,alb23}). 

The glass transition is continuous and not a genuine phase transition, although $T_g$ is fairly well defined for a given cooling rate, typically within 1\%. At $T_g$ the system falls out of metastable equilibrium because the time to reach equilibrium after an external disturbance, the so-called $\alpha$ relaxation time $\ta$, exceeds the laboratory time scale. By the fluctuation-dissipation theorem $\ta$ is also the characteristic time of the equilibrium fluctuations. This quantity is termed the Maxwell relaxation time. In the Maxwell model of viscoelasticity \cite{moo57,lam78,dyr06} $\ta$ is given by  

\be\label{eq:Max}
\ta=\frac{\eta}{G_\infty}
\ee
in which $G_\infty$ is the high-frequency plateau shear modulus corresponding to MHz frequencies and above (sometimes denoted by $G_p$). In this expression the temperature dependence of $G_\infty$ is insignificant, so upon cooling $\ta$ increases roughly proportionally to $\eta$. With $G_\infty\sim 10^9$ Pa the typical ordinary liquid viscosity $10^{-3}$ Pa$\cdot$s corresponds to $\ta\sim 10^{-12}$s, which is comparable to vibration (phonon) times. On the other hand, equating $\ta$ to the typical cooling time for producing a glass $\sim 10^3$ s leads to $\eta\sim 10^{12}$ Pa$\cdot$s.

A note on terminology: The term ``glass'' is used below whenever a highly viscous liquid is not in thermodynamic equilibrium, while ``liquid'' is reserved to a system in (metastable) equilibrium, i.e., one that is fully characterized by pressure and temperature with no memory of its past. The terms ``glass-forming liquid'' and ``ultraviscous liquid'' are used synonymously, reflecting the fact that once a liquid has been supercooled to the ultraviscous state by avoiding crystallization, glass formation is inevitable upon continued cooling.

\begin{figure}[H]
\begin{center}
\includegraphics[width=4.5 cm]{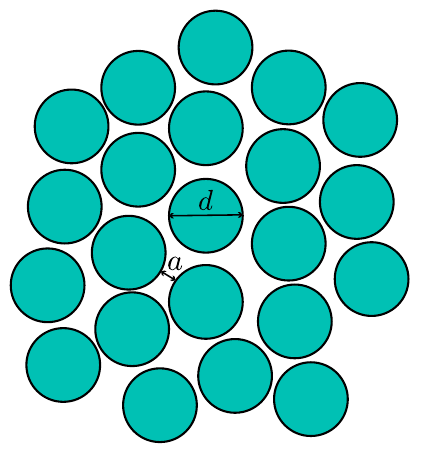}
\caption{\label{fig:2dliquid}
Hard-sphere liquid in two dimensions. There are frequent collisions because the particles almost touch. For a rough order-of-magnitude estimate of the system's transport properties, however, one can assume that $a\sim d$.}
\end{center}
\end{figure}

\noindent\textbf{Ordinary Liquids.}
Consider a pure substance above its melting temperature. As a crude approximation one may adopt the hard-sphere (HS) model consisting of identical particles that do not interact, except by never overlapping (\fig{fig:2dliquid}). Although this model is highly idealized, it is quite successful in reproducing the structure and dynamics of simple liquids \cite{han13,barrat}. Temperature plays no role in the HS model except for determining the average particle velocity, i.e., the relevant time scale. The only nontrivial thermodynamic variable is the density (packing fraction). The HS model may be regarded as a mathematician's idealized liquid; an alternative generic model in which temperature does play a role is the EXP system defined by the exponential repulsive pair potential \cite{bac14a,dyr16,EXPI}.

In the HS liquid each particle is close to several others (\fig{fig:2dliquid}), and the frequent particle collisions result in an erratic particle motion. This is different from what happens in the gas phase in which the mean-free path between collisions is much larger than the particle diameter. It is useful to discuss the physics of the HS liquid in terms of two diffusion coefficients, the particle-diffusion coefficient $\Dp$ and the transverse-momentum diffusion coefficient $\Dm$. The latter is the kinematic viscosity of the Navier-Stokes equation, $\Dm=\eta/\rho$ in which $\rho$ is the mass density \cite{lan59_fluid}, while $\Dp$ is defined from the long-time mean-square particle displacement via $\langle\Delta x^2(t)\rangle=2\Dp t$. 

Writing $A\sim B$ to indicate that $A$ and $B$ are within one or two decades of each other, the HS liquid is characterized by 

\be\label{eq:diff_ord}
\Dp\sim\Dm\,.
\ee
This applies not only for HS but for all ordinary liquids. $\Dp$ may be estimated by considering a random walk leading to $\Dp\sim d^2/\ta$. Typical experimental values of $\Dp$ and $\Dm$ are of order $10^{-7}\, {\rm m^2/s}$, which may be arrived at from $d\sim 10^{-10}$ m and $\ta\sim 10^{-13}$ s or from $\Dm=\eta/\rho$ with $\eta\sim 10^{-3}$ Pa$\cdot$s and $\rho\sim 10^3\, {\rm kg/m^3}$. 

When the viscosity of a glass-forming liquid upon cooling increases by many orders of magnitude, surprisingly little structural change takes place \cite{ber11}. This means that \fig{fig:2dliquid} is still a fairly good representation as regards structure, which raises the question: how should one think of a liquid with the extremely slow dynamics characterizing the approach to the glass transition?

\noindent\textbf{Ultraviscous Liquids.}
In thermal equilibrium, atoms/molecules have velocities proportional to the square root of temperature, but this motion does not necessarily imply lasting particle displacement. In a crystal, for instance, all thermal motion goes into vibrations around the equilibrium positions. One likewise expects effective particle motion in an ultraviscous liquid to be minute, because in order to move a particle with a certain velocity, a force is required that is proportional to the viscosity. Indeed, according to the Stokes-Einstein relation $\Dp$ is inversely proportional to the viscosity \cite{barrat,cos19}. Although derived by reference to macroscopic hydrodynamics, the Stokes-Einstein relation works well for simple liquids even on the molecular scale \cite{jsch}. The relation is violated by 1-3 orders of magnitude for liquids approaching the glass transition \cite{fuj92,tar95,sil99,edi00}, but this does not alter the fact that when viscosity increases upon cooling, $\Dp$ decreases roughly as much: $\Dp\propto\eta^{-x}$ with $x\geq 0.8$ \cite{swa03,har09,mal10a}. Thus when the viscosity -- and thereby $\Dm$ -- increases by a factor of $10^{15}$ by cooling from $T_m$ to $T_g$, $\Dp$ at the same time \textit{decreases} enormously. Interestingly, the heat-diffusion coefficient $\Dh$ changes only insignificantly upon cooling, even into the glassy state \cite{Prop_gas_liq,khr23}. To summarize, an ultraviscous liquid is characterized by

\be\label{eq:diff_extr}
\Dp\,\ll\,\Dh\,\ll\,\Dm\,.
\ee

\noindent\textbf{Flow Events.}
Since effective particle motion is exceedingly slow in an ultraviscous liquid while velocities are not, most motion must go into vibrations. Two possible scenarios can realize this. The vibrations can take place around average positions that change continuously, but exceedingly slowly. Alternatively, sudden rare localized ``flow events'' rearrange a handful of particles. Experiments on colloidal \cite{hun12}, molecular \cite{sto22}, and metallic \cite{wan19} glass-forming liquids, as well as computer simulations \cite{sch00a}, favor the latter scenario. This does not mean that very slow position changes are absent \cite{wid09}; they do take place and are important for the physics. According to the \stf picture as detailed below, however, these minor collective displacements are an \textit{effect} of flow events taking place in a solid-like structure.

It is an old idea that particle motion in a glass-forming liquid proceeds via flow events. In his seminal 1948 review Kauzmann referred to flow events as ``jumps of molecular units of flow between different positions of equilibrium in the liquid's quasicrystalline lattice'' \cite{kau48}. Mooney in 1957 poetically referred to a flow event as ``a quantum of liquid flow'' \cite{moo57}, and many subsequent papers have embraced this picture of viscous liquid dynamics \cite{ada65,gol72,dyr87,lub07,mir14b,wan19a}. The physics, of course, lies in what determines the energy barriers of flow events, how these events correlate in space and time, and how they control physical properties.

\begin{figure}[H]\begin{center}
		\includegraphics[width=4.5 cm]{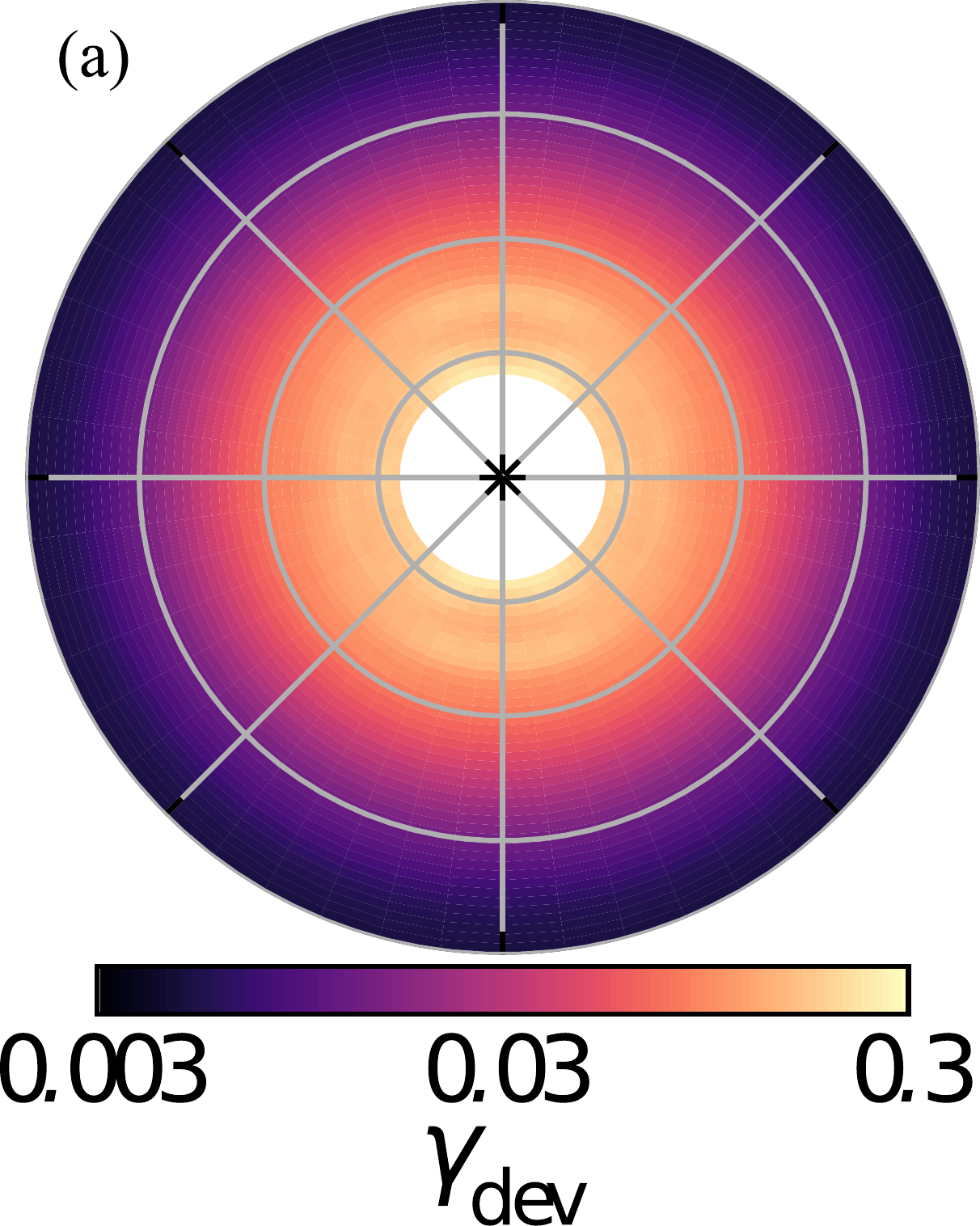}
		\includegraphics[width=7 cm]{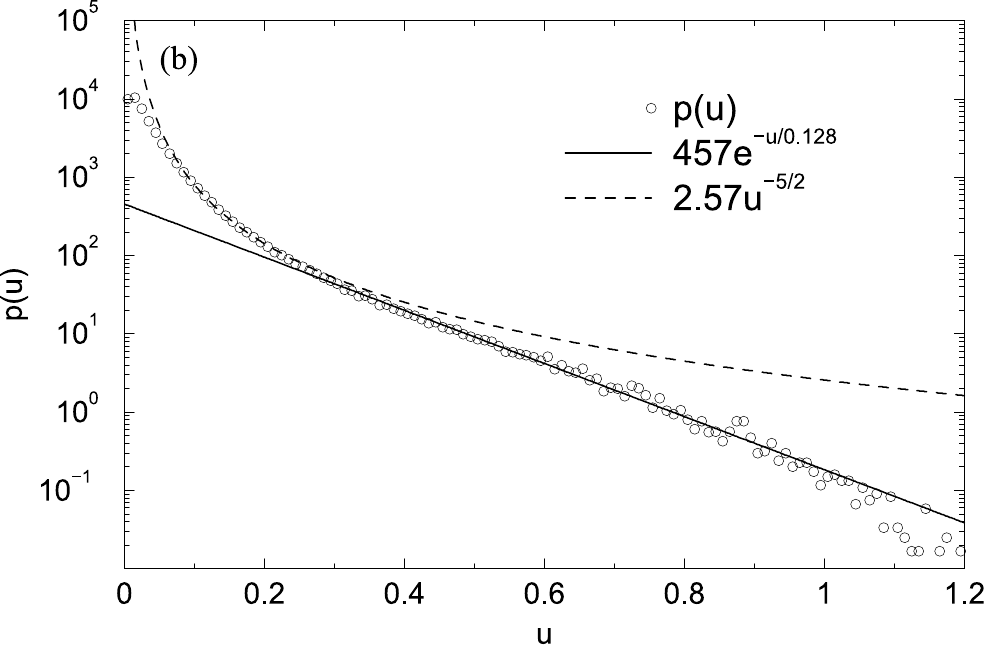}     
	\end{center}
	\caption{Displacements induced by a flow event in a glass-forming liquid. \label{fig:flow}
		(a) Average deviatoric strain displacement in the surroundings of flow events in simulations of a 2d polydisperse glass-forming liquid (the radial axis is logarithmic with two units corresponding to a factor of ten). The data conform to the long-range decay $\propto 1/r^2$ predicted by solid-state elasticity theory. 
		Reproduced from Ref. \onlinecite{cha21}. Copyright 2021 Authors, licensed under a Creative Commons License.        
		(b) Displacement probability, $p(u)$, in the surroundings of flow events of a 3d binary glass-forming liquid. The vast majority of displacements are small and conform to $p(u)\propto u^{-5/2}$ (dashed curve), which is a consequence of the solid-state elasticity-theory prediction $u\propto 1/r^2$ for $r\to\infty$ \cite{dyr99,sch00a}. The limited number of particles of this early simulation ($N=500$) accounts for the deviation from the dashed curve at small $u$. 
		Reproduced from Ref. \onlinecite{sch00a}. Copyright 2000 AIP Publishing. }
\end{figure}

Why are flow events rare in an ultraviscous liquid? This was reflected upon by Goldstein in his famous 1969 paper \cite{gol69}. He identified Kauzmann's ``positions of equilibrium'' with minima of the potential-energy function. Flow events are rare because the barriers to be overcome going from one minimum to another are much larger than $k_BT$. Potential-energy minima are nowadays referred to as ``inherent states'' \cite{sti95}. Goldstein's picture is that the dynamics of an ultraviscous liquid involves numerous vibrations around an inherent state. These vibrations do not contribute to the slow dynamics and may be eliminated by focusing on the ``inherent dynamics'' defined as the time sequence of inherent states \cite{sch00a,dol03a}.

Anticipating that an ultraviscous liquid is like a solid on short length scales, any flow event leads to minor deformations in its surroundings \cite{cha13,lem14,ill16}. These may be detected, e.g., by NMR experiments \cite{boh98}. Sufficiently far from a flow event linear elasticity applies, and in 3d the induced particle motion scales as $1/r^2$ for $r\to\infty$ where $r$ is the distance to the flow event \cite{lan59_elast} (\fig{fig:flow}). To show this one uses the mechanical-equilibrium requirement that the time-averaged force on each particle is zero both before and after a flow event. A single flow event's effects on the surroundings may be reproduced by imagining external forces acting on a small surface surrounding it \cite{esh57,dyr99a,lem15}. By momentum conservation, each of these forces leads to a momentum flow into the surroundings $\propto 1/r^2$ for $r\to\infty$. Since the forces sum to zero, this implies an overall momentum flow (stress tensor) that is the spatial derivative, i.e., $\propto 1/r^3$. According to elasticity theory \cite{lan59_elast}, the stress tensor change is linearly related to the strain field, which is formed from derivatives of the displacement field that consequently must scale as $\propto 1/r^2$ \cite{lem18}. In 2d the stress and strain fields induced by a flow event scale as $\propto 1/r^2$ for $r\to\infty$ and the particle displacements as $\propto 1/r$.

\noindent\textbf{Solidity Length.}
The above and \fig{fig:flow} suggests that the physics of an ultraviscous liquid is reminiscent to that of a solid. Real-life solids are mostly crystalline with grain boundaries separating small crystals containing point defects. In thermal equilibrium, however, the solid state of any pure substance is a single crystal with no line defects or grain boundaries, while a few point defects like vacancies and interstitials, i.e., missing or extra atoms, are present \cite{ash76}. Point defects can jump to neighboring positions by overcoming a barrier much larger than $k_BT$, just like flow events in ultraviscous liquids.

The effect of a flow event on its surroundings is not instantaneous due to the finite sound velocity. After a flow event, others take place nearby that likewise send out spherical ``waves'' of minor adjustments of the particle positions. Far from the original flow event, the adjustments originating from many other flow events interfere with and increasingly smear out the effect of the original flow event. We proceed to estimate the length scale beyond which this significantly dampens the effects of a flow event, which defines the system's ``solidity length'' $\ls$. The system is solid-like on length scales smaller than $\ls$, but not on larger length scales. A related concept is the ``shear penetration depth'' quantifying how far an external shear disturbance penetrates into the liquid \cite{wei16a}.

\begin{figure}[H]
\begin{center}
\includegraphics[width=6 cm]{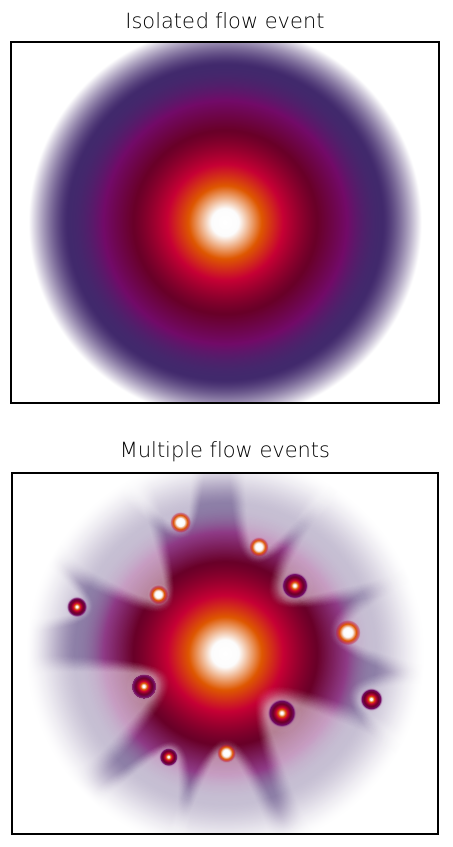}
\caption{ \label{fig:fe} An isolated flow event and its screening by subsequent nearby flow events (artist's impression). 
The radius of the sphere inside which the effects of the original flow event are felt in full defines the solidity length $\ls$.}
\end{center}
\end{figure}

To estimate $\ls$ we make the crude approximation that the average time between two flow events involving a given molecule is $\ta$. If $a$ is the average intermolecular distance, a sphere of radius $\ls$ contains or order $N \cong (\ls/a)^3$ sites for potential flow events. Flow events are not independent and uncorrelated, but for simplicity we ignore this and estimate the average time between two flow events within the sphere to be $\ta/N$. The solidity length is determined by requiring that this equals the time it takes a sound wave to travel $\ls$, which is $\ls/c$ where $c$ is the sound velocity \cite{dyr99}, leading to

\be\label{eq:ls}
\ls^4
\,\cong\,c\,a^3\,\ta\,.
\ee
For an ordinary liquid the derivation makes little sense, but if one nevertheless substitutes $c\sim 10^3$ m/s, $a\sim 10^{-10}$ m, and $\ta\sim 10^{-13}$ s into \eq{eq:ls}, the result is $\ls\sim 10^{-10}$ m. For a liquid approaching the glass transition, $\ta\sim 10^2$ s leads to $\ls\sim 10^{-6}$ m. This length is much larger than those discussed in connection with dynamic heterogeneities of glass-forming liquids \cite{dynhet,kar14,wya17,bir23}. Note that the derivation of \eq{eq:ls} is general and applies also, e.g., for network-forming liquids like silica. 

A single crystal with point defects also has a finite solidity length, but here the concept is not relevant because the crystal structure defines solid-like particle correlations over distances stretching to infinity. For an ultraviscous liquid, on the other hand, due to the lack of a lattice, rigid distance correlations apply only below the solidity length. 

Furukawa has argued that the length $\xi$ defined by $\xi^4\equiv a^4\ta/\tau_0$, in which $\tau_0$ is a microscopic time, is the characteristic length over which long-lived stress is sustained \cite{fur13,fur19}. He proposed that an ultraviscous liquid may be regarded as an ordinary liquid composed of clusters of size $\xi\sim\ls$, and that hydrodynamics only applies on length scales above $\xi$. Much of the physics probed in experiments actually takes place \textit{below} the solidity length, e.g., in dielectric relaxation or NMR experiments probing a molecular average property. In fact, measuring a macroscopic dynamic property like the viscosity $\eta$ becomes increasingly difficult as the glass transition is approached \cite{hec13}.

\noindent\textbf{Conservation Laws.}
Below the solidity length the laws of conservation of the number of particles, momentum, and energy play roles different from in ordinary liquids where these laws form the basis of hydrodynamics \cite{kad63,mar72,han13,jsch,bag22}. Consider first particle conservation. A molecular dynamics simulation keeps track of the individual particles, but things look different in a coarse-grained description based on a continuous density field $\rho(\br,t)$. This field is constant in time until it changes due to a flow event, a change that below the solidity length may be regarded as instantaneous. In general, flow events are not spherically symmetric, but this assumption can be made initially when discussing the density change at the flow event center. 

A spherically symmetric flow event leads as mentioned to purely radial displacements in the surroundings $\propto 1/r^{2}$ \cite{lan59_elast}. The divergence of the displacement field, which determines the local density change, is zero (compare Gauss' law for the point-charge electric field $\propto 1/r^2$). On the other hand, this radial displacement can only take place if there is a density change at the flow-event center. Hence, if the density change after coarse-graining over a few molecular distances is denoted by $\Delta\rho(\br)$, a flow event at $\br_0$ leads to

\be\label{eq:denschan}
\Delta\rho(\br)=0\,
\ee
in its surroundings. Comparing the situation before and after the flow event, local particle conservation will appear to be violated because the density changes only at $\br_0$. What happens is reminiscent of Hilbert's hotel, the full infinite hotel that hosts new arrivals by asking all guests to move to a room of one higher number. 

Below the solidity length flow events may as mentioned be regarded as instantaneous. This leads to a coarse-grained description with no visible trace of particle-number conservation \cite{dyr06a},

\be\label{eq:dens_dyn}
\dot\rho(\br,t)
\,=\,\sum_j b_j\delta(\br-\br_j)\delta(t-t_j)\,.
\ee
Here $b_j$ is a dimensionless measure of the magnitude of the flow event at time $t_j$ and position $\br_j$. Apparent density non-conservation holds also if one takes into account the minor density changes of the more realistic anisotropic Eshelby-type flow events discussed, e.g., in Refs. \onlinecite{esh57,esh59,dyr99a,lem14}. In that case, the right-hand side of \eq{eq:dens_dyn} acquires an additional ``advective'' term reflecting the long-ranged minor effects of flow events on the density, a term that also after coarse-graining will conform to density conservation.

Below the solidity length mechanical equilibrium applies in the time between flow events, i.e., time-averaged forces are zero. In a coarse-grained description this is expressed as zero divergence of the stress tensor, $\str(\br,t)$ \cite{lan59_elast}

\be\label{eq:zerodiv}
\partial_\mu \str(\br,t)
\,=\,0
\ee
in which $\br=(x_1,x_2,x_3)$ and $\partial_\mu$ is the spatial derivative with respect to $x_\mu$. In this approach the dynamics is regarded as a series of instantaneous transitions between states of mechanical equilibrium, each of which conform to \eq{eq:zerodiv}. 

Momentum conservation likewise plays little role in the dynamics below the solidity length. The situation is similar to that of point-defect motion in a crystal for which one would not invoke momentum conservation to explain the physics. The same applies for energy conservation: in an ultraviscous liquid, energy flow predominantly takes place via heat conduction, just like in a solid, and this process is irrelevant for the rate of flow events or for explaining how these correlate in space and time. We are not suggesting that strict particle, momentum, or energy conservation is violated, merely that these conservation laws are not relevant for understanding the physics of glass-forming liquids \cite{dyr07c,dyr07a}.

\begin{figure}[H]
\begin{center}
        \includegraphics[width=8 cm]{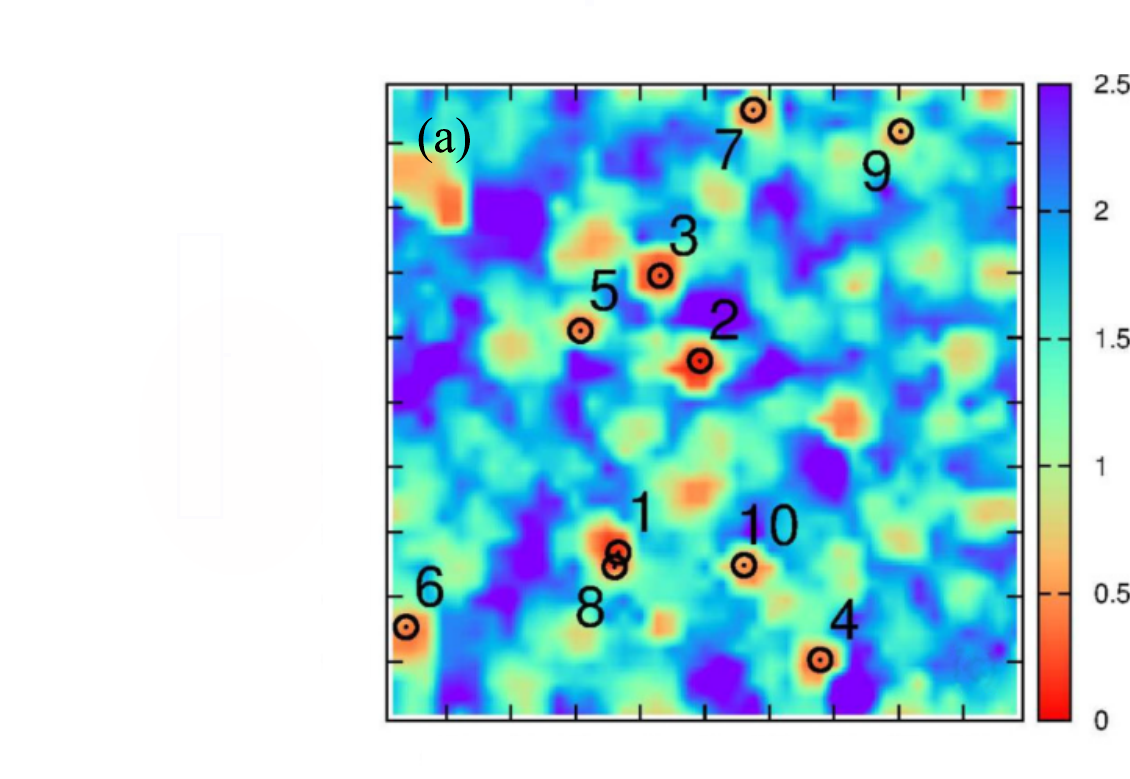}
        \includegraphics[width=5.5 cm]{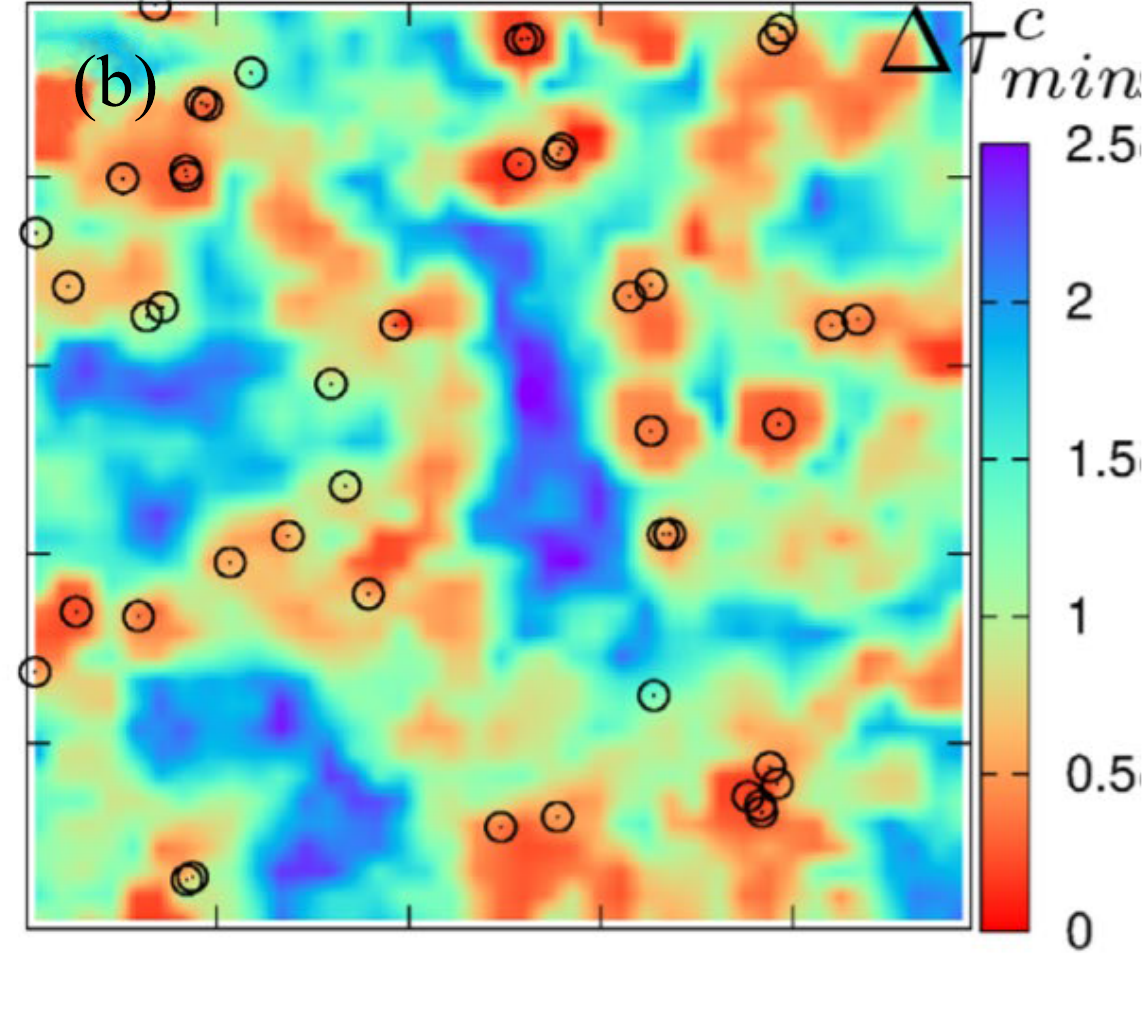}
\end{center}
        \caption{\label{fig:soft} Soft spots in 2d systems.
        (a) Softness probed as the local yield stress of a binary Lennard-Jones glass in which red and blue mark soft and hard regions, respectively. The numbered points give the time sequence of plastic flow events when the system is sheared at zero temperature, which are clearly located at the soft spots. 
        Reproduced from Ref. \onlinecite{bar18}. Copyright 2018 American Physical Society.
        (b) Analogous results for an equilibrium ultraviscous 2d liquid in which softness is probed by the local yield stress of the system's inherent state, i.e., potential-energy minimum, at a specific time. The circles mark the first 50 subsequent flow events. Like the plastic flow events of the glass in (a), these preferably take place at soft spots. 
        Reproduced from Ref. \onlinecite{ler22}. Copyright 2022 American Physical Society.}
\end{figure}

\noindent\textbf{Flow Events in Plastic Flow of Glasses.}
In recent years, the understanding of glass-forming liquids has benefited
greatly by learning from studies of forced flow of glasses. When a glass is subjected to a gradual shear deformation, it eventually yields by deforming irreversibly \cite{sch07a,lin14a,voi14,huf16,nic18}. The last 15 years has brought tremendous progress in the understanding of zero-temperature plastic flow of glasses \cite{ric20}, which proceeds as a sequence of sudden, localized flow events \cite{fal98,tan06,bou07}. These do not take place at random locations, but at soft spots in the glass, compare \fig{fig:soft}(a) \cite{ler16}.

Different methods have been suggested for identifying soft spots \cite{ric20}. An early approach was to look for particles with a large finite-temperature vibrational mean-square displacement, a simple indicator that the potential is soft at the particle in question \cite{wid06}. The mean local potential-energy fluctuation has also been used as a soft-spot identifier \cite{zyl17}. The local-yield-stress method \cite{pat16,bar18} considers a small sphere and constrains the outside system to deform affinely such that only the atoms inside the sphere can relax when the system is shear deformed. Different sphere locations and possible shear deformations are probed to identify the position of smallest local yield stress. This method works well for identifying the sites of plastic rearrangement, but requires knowledge of the interaction potentials. Methods for identifying sites for plastic flow events based purely on structural information have also been devised \cite{cos11,roy15} using, e.g., a mean-field caging potential \cite{nan21} or machine-learning techniques \cite{cub17,boa20}. One such method \cite{cub17} defines ``softness'' as a weighted integral over local radial pair-correlation functions and optimizes the weights for predicting sites of rearrangement by learning from plastic flows. The results obtained correlate well with those of the yield-stress method \cite{cub17,boa20}.

An alternative approach utilizes the fact that soft spots give rise to localized phonon modes, implying that the latter are good predictors for plasticity \cite{wid08,tan10,man11,ler21,wu23}. Low-energy localized vibrational modes may be identified by a third-order expansion of the potential-energy function, a clever method that avoids the hybridization with the low-frequency sound-wave modes found by diagonalizing the Hessian \cite{kap20}. -- Despite the above quoted strong indications of a connection between soft spots and relaxation sites, this link has exceptions and is not universally agreed upon \cite{jac14,miz20,li22}.

To summarize, plastic flow takes place via sudden flow events. The physics is similar to what happens in an equilibrium ultraviscous liquid, in which flow events are also located at ``soft spots'' (\fig{fig:soft}(b)). A difference is that the time-sequence of flow events of a zero-temperature plastic flow is deterministic, while a glass-forming liquid's flow events are stochastic. Another difference is the lack of isotropy of a plastic flow, which leads to preferred orientations of the Eshelby stress fields, to which we now turn.

\noindent\textbf{Strain and Stress Correlations in the Liquid.}
We return below to the idea that flow events are controlled by the system's elastic properties and focus here on another property of glasses, the fact that a flow event generally induces a quadrupolar stress-field change in the surroundings \cite{pic04}. This is explained by the 1957 theory of solid inclusions by Eshelby \cite{esh57}, which applies also to disordered solids because these are effectively homogeneous on a long length scale. Eshelby calculated the long-ranged stress and strain changes of an inclusion by replacing it with localized forces in an elastic continuum. Each force gives rise to a momentum current into the solid, and since these forces must sum to zero, the result is a quadrupolar stress field \cite{esh57,esh59}.

\begin{figure}[H]\begin{center}
       \includegraphics[width=7.4 cm]{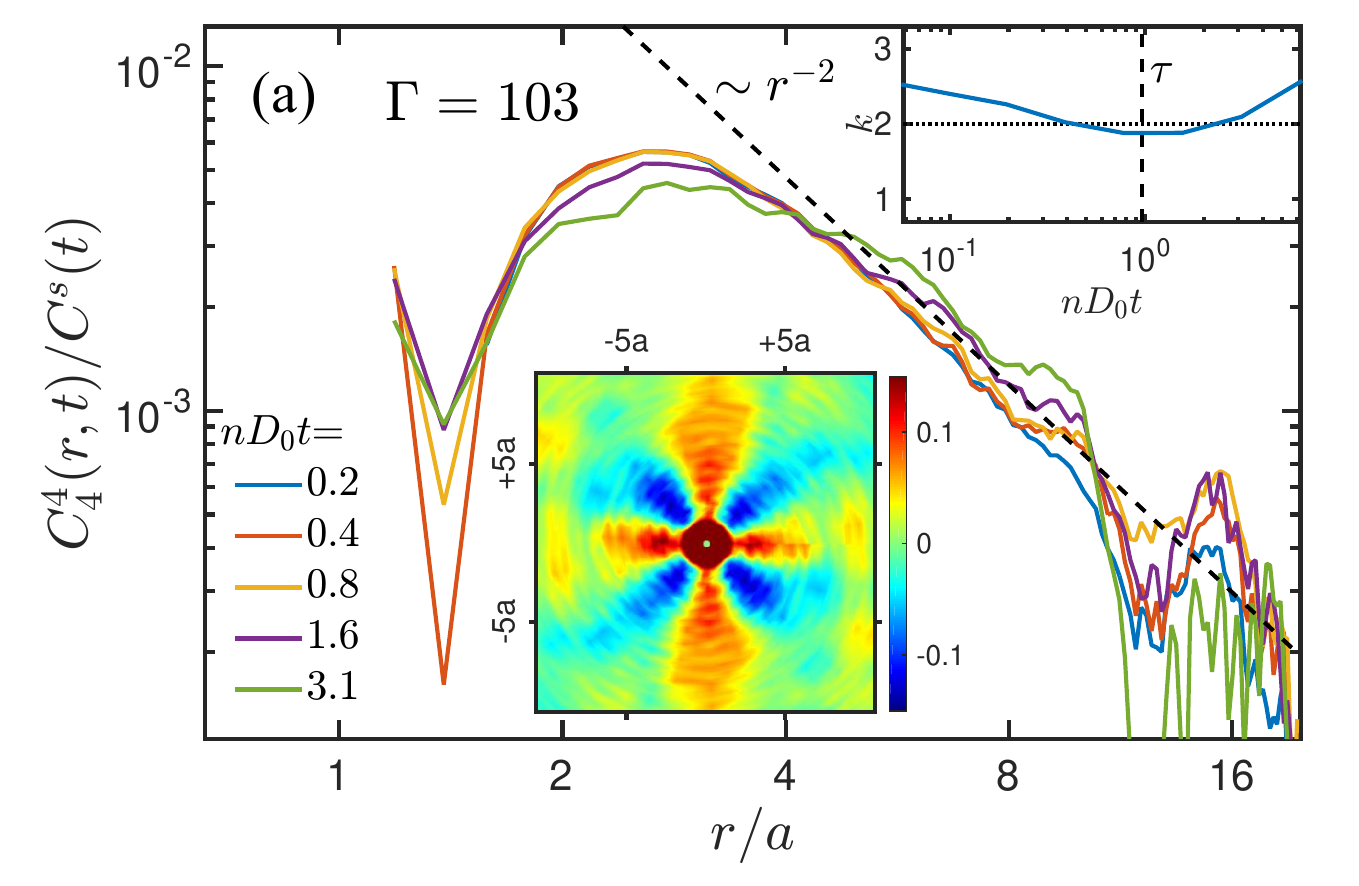}
       \includegraphics[width=4.4 cm]{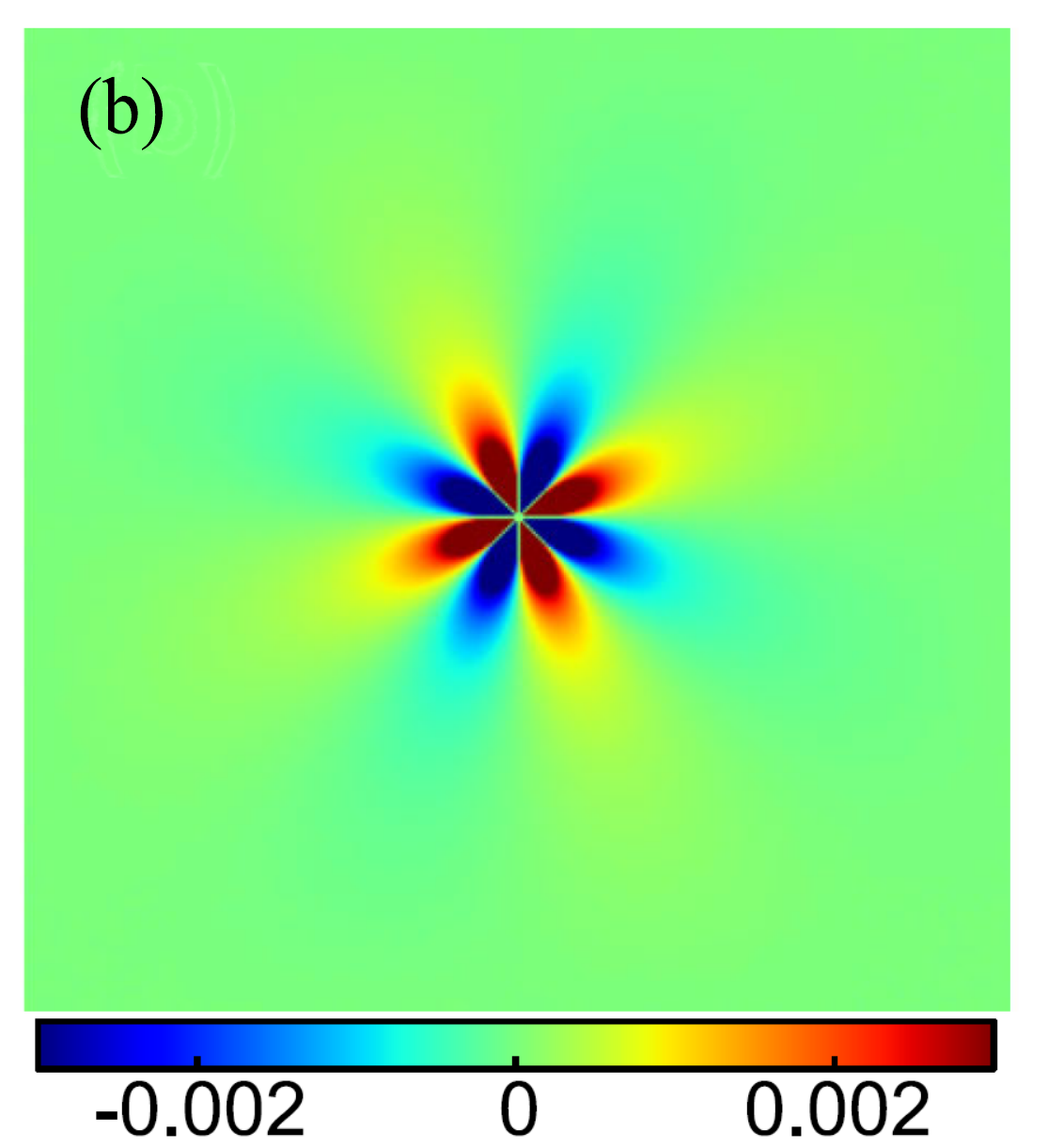}      
       \end{center}
        \caption{\label{fig:stress} Spatial strain and stress correlations in ultraviscous liquids. 
        (a) Experimental data for the strain correlations of a 2d colloidal glass-forming liquid. The lower inset shows the spatial correlation function of the $xy$ strain-tensor change over a time $t\gg\ta$ where green is zero. The curves are normalized spherical-harmonics projections of this function at different times, which are proportional to $1/r^2$ as predicted for Eshelby patterns in 2d \cite{esh57,esh59,lan59_elast,lem14,lem18}.
        Reproduced from Ref. \onlinecite{ill16}. Copyright 2016 American Physical Society.
        (b) Stress-tensor correlations in a 2d binary viscous liquid. The figure shows the correlations between the normal-shear-stress increment of a single flow event. Reproduced from Ref. \onlinecite{lem14}. Copyright 2014 American Physical Society.}
\end{figure}

The obvious question is whether the long-ranged stress correlations observed in glasses \cite{chi11} exist also in glass-forming liquids \cite{cho16,buc18,has21,li22}. One expects this to be the case below the solidity length because here the properties of the liquid's inherent states -- each of which corresponds to a $T=0$ glass -- is inherited by the equilibrium liquid (\fig{fig:stress}) \cite{fle15,cho16,mai17,klo22,ste22}.

Lemaitre worked out the theory for the stress-tensor spatial autocorrelation function in a disordered isotropic solid. Remarkably, the $3\times 3\times 3\times 3$ tensor $\langle\sigma_{\alpha\beta}(\br)\sigma_{\gamma\delta}(\br')\rangle$ is determined by just two functions of $|\br-\br'|$ \cite{lem14,lem18}. In general, if $X$ is the space or space-time coordinate and $D_X$ is a linear differential operator of a field theory with equation of motion ``$D_X\phi(X)=$ Noise'', one has $D_{X'}\langle\phi(X)\phi(X')\rangle=0$ whenever $X'$ differs from $X$ in all coordinates. Thus according to the \stf picture, as a function of $\br'$ the autocorrelation function $\langle\sigma_{\alpha\beta}(\br)\sigma_{\gamma\delta}(\br')\rangle$ obeys \eq{eq:zerodiv}. In particular, it conforms to the Eshelby theory \cite{esh59} for $|\br-\br'|\to\infty$. The applies to the more general space-time autocorrelation function $\langle\sigma_{\alpha\beta}(\br,t)\sigma_{\gamma\delta}(\br',t')\rangle$.

Long-ranged stress and strain correlations only exist below the solidity length, however. At longer length scales the effects of one flow event are smeared out by those of others. Thus beyond the solidity length $\ls$, an exponential decay of the spatial stress autocorrelation function is to be expected. This means that in the liquid phase, the Lemaitre spatial autocorrelation functions \cite{lem18} should be multiplied by a factor $\sim\exp(-|\br-\br'|/\ls)$. 

We finally note that the non-zero stresses of an ultraviscous liquid modify nearby flow-event energy barriers. This is not taken into account in attempts to identify likely positions of flow events from specific structures \cite{roy15}, which may explain why such attempts have only been moderately successful \cite{wei19}.

We end this perspective by giving three examples of how the \stf picture elucidates experimental facts of glass-forming liquids.

\noindent\textbf{Dynamic Heterogeneity and Elastic Facilitation.}
An important finding of the 1990s was that the dynamics of a glass-forming liquid is spatially inhomogeneous \cite{hur95,sch96,sil99,edi00,glo00}. At any given time, there are regions of considerable molecular motion and regions of little \cite{tan19,ber21}. This provides a simple explanation of the observed violation of the Stokes-Einstein relation between viscosity and diffusion coefficient \cite{tar95,die97,edi00}: Fast particles take advantage of rapidly relaxing regions and contribute a lot to $\Dp$, but little to the overall structural relaxation rate quantified by $\ta$ via the viscosity.

\begin{figure}[H]\begin{center}
       \includegraphics[width=5 cm]{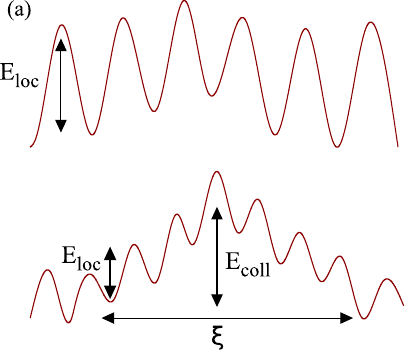}
       \includegraphics[width=5 cm]{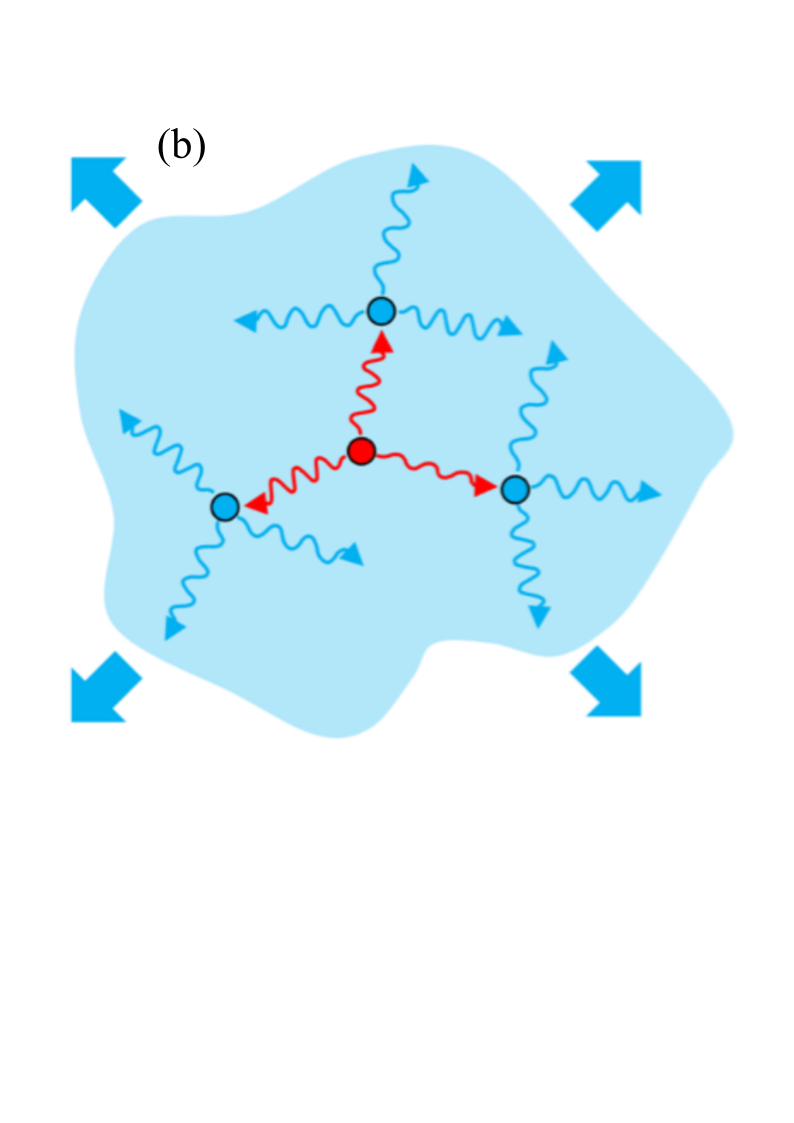}    
       \includegraphics[width=6 cm]{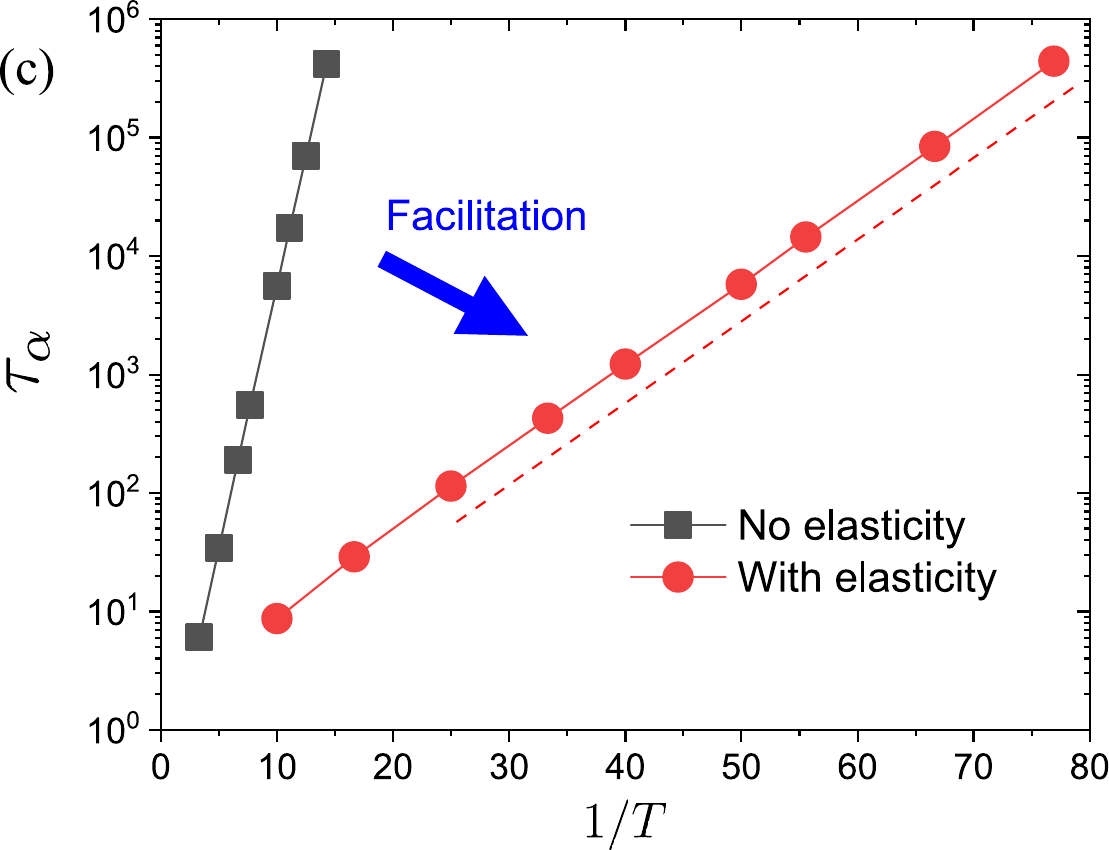}
       \end{center}
        \caption{\label{fig:facilitation}Elastic facilitation. 
        (a) Two schematic free-energy landscapes in configuration space. The dynamics can be dominated by local energy barriers (upper panel) or by the growth of cooperative effects over some distance $\xi$ (lower panel). 
        Reproduced from Ref. \onlinecite{cia23}. Copyright 2023 Authors, licensed under a Creative Commons License.
        (b) Avalanche of flow events induced by a single flow event that lowers some barriers in the surroundings, thereby catalyzing other flow events that induce yet others, etc. 
        Reproduced from Ref. \onlinecite{tah23}. Copyright 2023 Authors, licensed under a Creative Commons License.
        (c) Temperature dependence of $\ta$ in a simple model with and without elastic facilitation. 
        Reproduced from Ref. \onlinecite{oza23}. Copyright 2023 American Physical Society. }
\end{figure}

What controls the temperature dependence of $\ta$? \Fig{fig:facilitation}(a) presents two fundamentally different scenarios \cite{cia23}. In one case (upper panel), the local energy barrier controls the dynamics in the sense that it determines the overall relaxation rate. Alternatively, structural relaxation is a highly cooperative process that involves an entire sequence of flow events (lower panel). The latter scenario has been used to explain dynamic heterogeneities and is predicted, e.g., by the random first-order transition (RFOT) theory \cite{wol12,bir23}. RFOT is inspired by the theory of spin glasses, which are systems with no elastic interactions. In RFOT the increase of the activation energy of $\ta$ upon cooling results from a correlation length $\xi$ that grows due to the decrease of entropy \cite{ada65,lub07}. The fundamental RFOT prediction, which has been proven rigorously for infinite dimensions \cite{parisibog}, is that thermodynamics control the dynamics. This is challenged, however, by the fact that swap algorithms in finite dimensions \cite{san00,nin17} can speed up computer simulations significantly without affecting the thermodynamics \cite{wya17,gav23}.

The local-barrier picture is the obvious one from the \stf view in which the situation is analogous to that of a plastic flow with flow events occurring at soft sites. That local barriers control the relaxation was demonstrated recently in simulations of a 3d polydisperse soft-repulsive-potential model \cite{cia23}. By systematically identifying the flow events starting at a given inherent state, it was shown that upon cooling the activation energy increases enough to account for the super-Arrhenius temperature dependence of $\ta$. This means that, at least for the model in question, the dynamics is not cooperative; in particular, no divergence of the relaxation time is predicted at a finite temperature \cite{san00,hec08, mck15}.

If $\ta$ is controlled by the individual flow-event activation energies, how does one explain dynamic heterogeneity? A promising candidate is 
facilitation, the idea that one flow event makes nearby flow events more likely \cite{fre84,gar02,rit03}. ``Elastic facilitation'' is a consequence of the fact that all flow events lead to long-ranged stress-tensor changes within the sphere defined by the solidity length. These lower the barriers of some potential nearby flow events and increase others \cite{lem14,cha21,zha22a,oza23}. Because the exponential function is convex, the net effect is that one flow event makes neighboring flow events more likely. Borrowing a term from NMR theory, this general mechanism has been referred to as ``rate exchange'' \cite{sch91a,die97,sil99}.

Elastic facilitation, which was first studied in glasses \cite{lem14,dua23}, is illustrated in \fig{fig:facilitation}(b). In a simple model \cite{oza23} this leads to a substantial reduction in the activation energy of $\ta$, \fig{fig:facilitation}(c). In Ref. \onlinecite{tah23} it was proposed that an entire avalanche of flow events may be triggered by a single one, similar to what happens in plastic flow of glasses \cite{nic18,sha20,ric23}. Reference \onlinecite{sca21} considered a simple facilitated trap model \cite{dyr87} and showed that it results in an asymmetric $\alpha$ loss peak with an excess wing (see also Refs. \onlinecite{has23,rid23}).

\noindent\textbf{Double-Percolation Scenario for Linear-Response Functions.}
Any linear response property is quantified by a complex frequency-dependent response function, $\chi(\omega)=\chi'(\omega)+i\chi''(\omega)$. It is sometimes stated that a major mystery of glass-forming liquids is the observed broad loss peaks, $\chi''(\omega)$. Certainly, a Debye loss peak, $\chi''(\omega)\propto \omega\tau/[1+(\omega\tau)^2]$ (which according to the fluctuation-dissipation theorem \cite{reichl} corresponds to an exponential time-autocorrelation function) is rarely observed. But one could also argue that the loss peaks are, in fact, surprisingly narrow. In particular, in the vast majority of glass-forming liquids dielectric, mechanical, and specific-heat loss peaks follow the Debye prediction on the low-frequency side. This striking fact implies the existence of a quite sharp long-time cut-off in the relaxation-time distribution $p(\tau)$ defined by $\chi''(\omega)= \int_0^\infty\omega\tau/[1+(\omega\tau)^2]p(\tau)d\tau$.  How can one understand this?

\begin{figure}[H]
\begin{center}
        \includegraphics[width=6 cm]{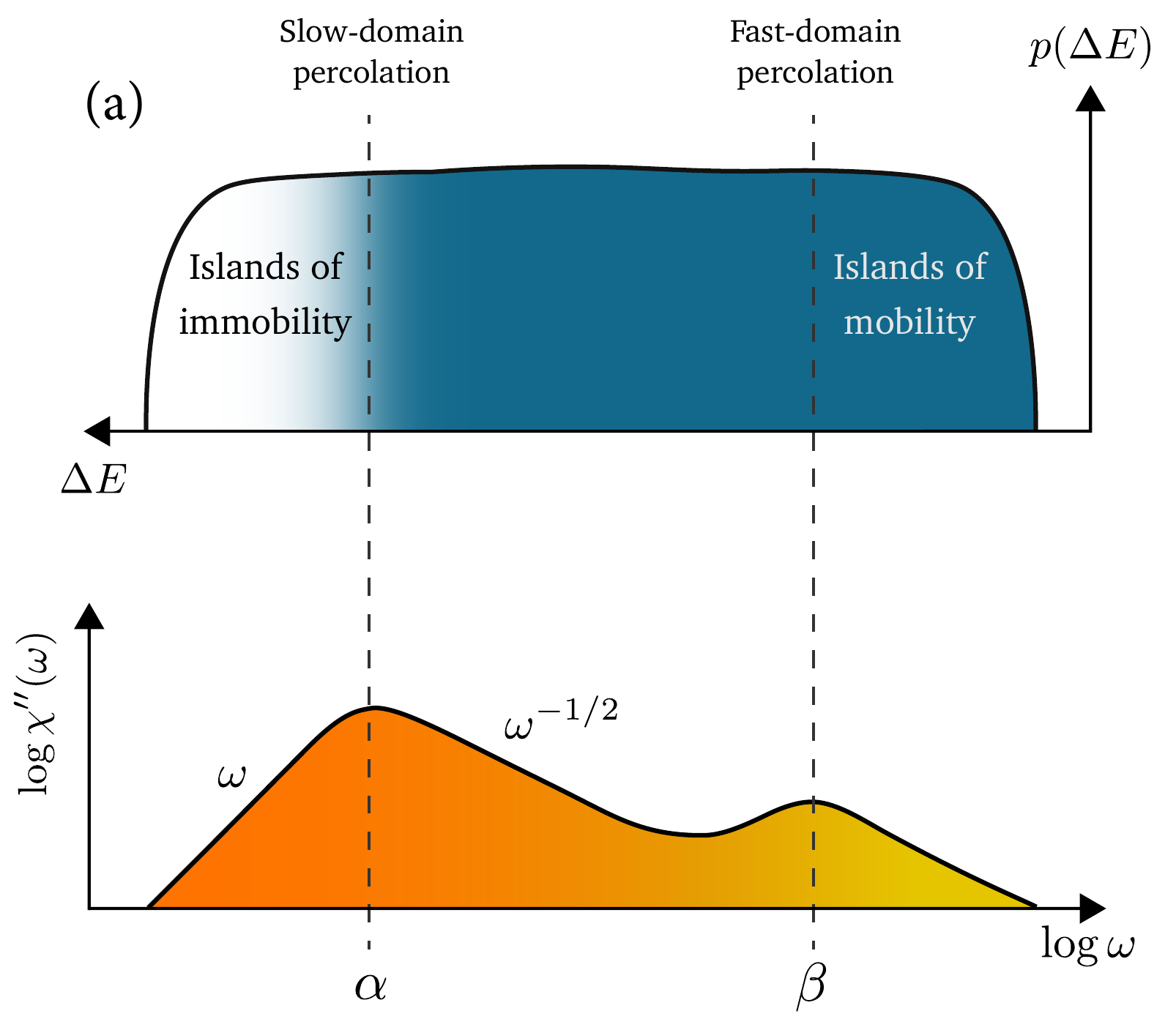}
        \includegraphics[width=5.5 cm]{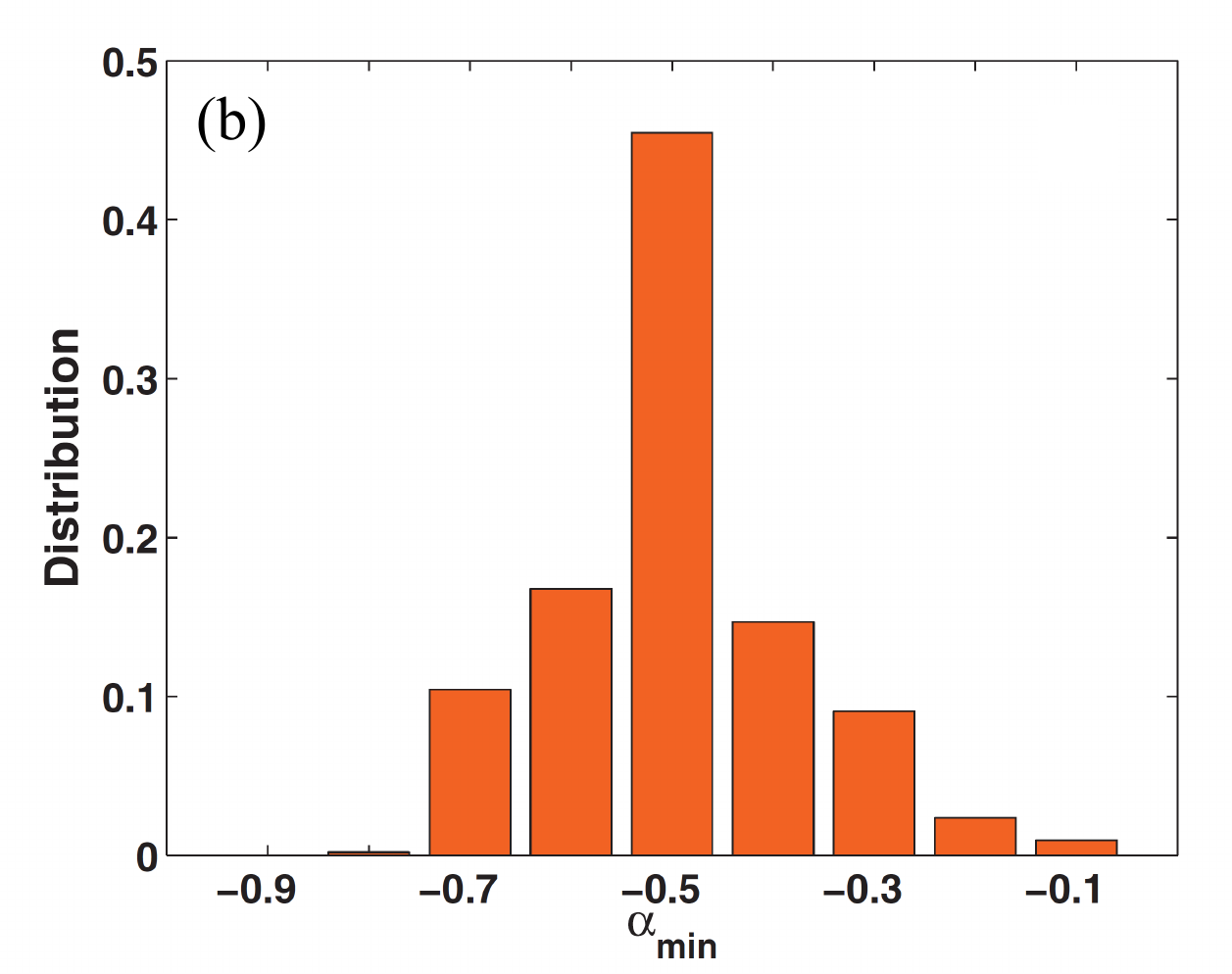}
        \includegraphics[width=6 cm]{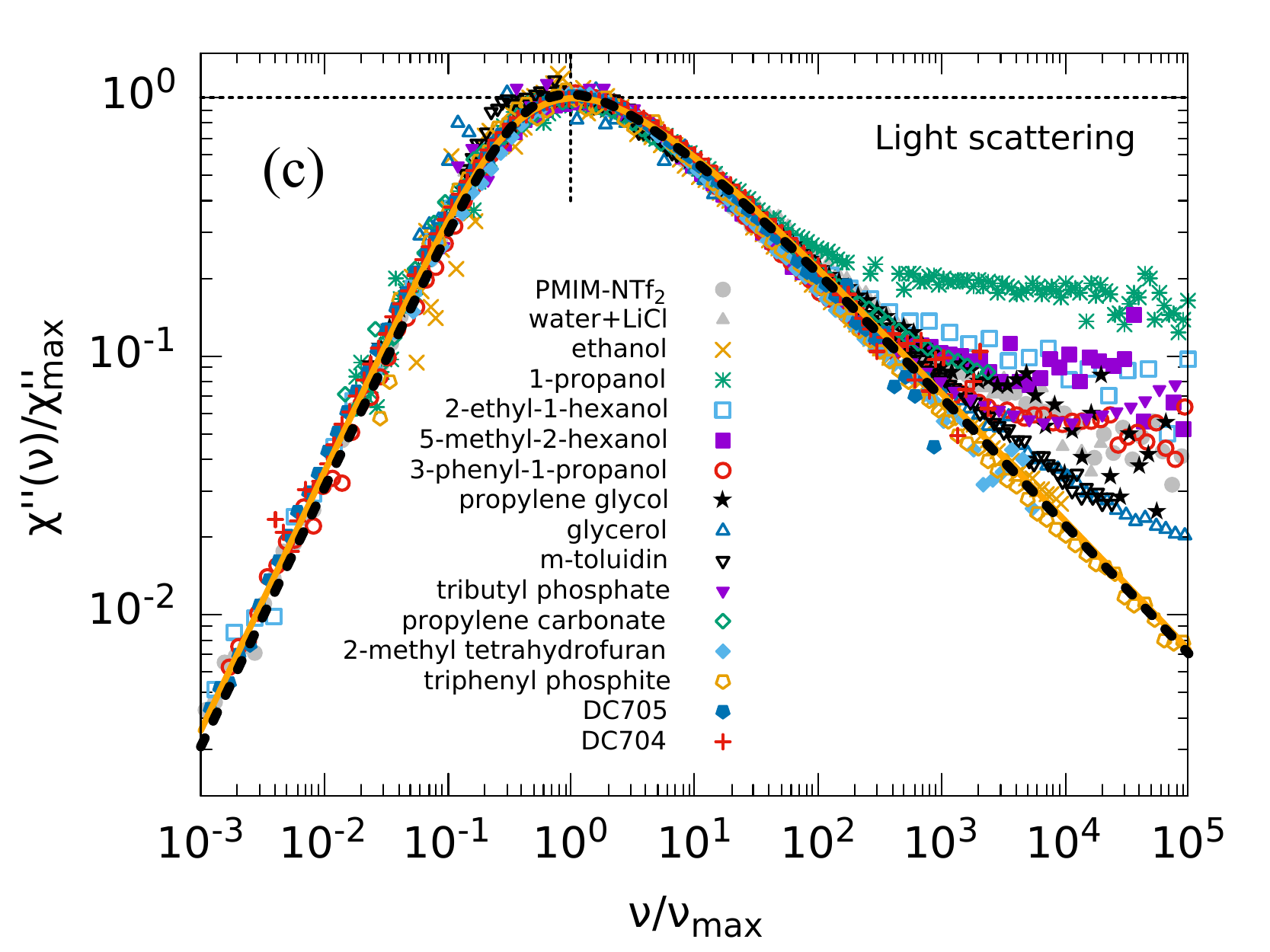}
        \caption{\label{fig:alfabeta} Linear-response consequences of the solidity of glass-forming liquids.
        (a) Double-percolation scenario for the frequency-dependent imaginary-part linear response, $\chi''(\omega)$, in the case of a very wide flow-event activation-energy distribution, $p(\Delta E)$. The distribution is constant in time, but the activation energy of any region changes over the $\ta$ time scale. The $\beta$ relaxation is marked by the frequency corresponding to where the islands of mobility percolate; the $\alpha$ relaxation is where the islands of immobility percolate. The islands of immobility do not contribute to any relaxation because their barriers are too high and await being lowered by elastic facilitation and/or by ``dissolving'' the slow-domain percolation cluster defining the $\ta$ time scale.
        (b) Histogram of the minimum slopes above the $\alpha$ dielectric log-log loss peaks for 347 spectra of 53 glass-forming liquids. The prevalent minimum slopes are close to $-0.5$. 
        Reproduced from Ref. \onlinecite{nie09}. Copyright 2009 Authors, licensed under a Creative Commons License.
        (c) Results from dynamic light scattering showing a $\chi''(\omega)\propto\omega^{-1/2}$ high-frequency decay for chemically quite different glass-forming liquids. The black dashed line is the imaginary part of $\chi(\omega)\propto 1/\sqrt{1+i\omega\ta}+1/(\sqrt{2}+\sqrt{1+i\omega\ta})$, which can be derived from the density-dispersion relation $\Gamma(k)=\Gamma_0+\Dc k^2$ that incorporates the apparent violation of density conservation via the $\Gamma_0 $ term \cite{dyr05}. 
        Reproduced from Ref. \onlinecite{pab21}. Copyright 2021 American Chemical Society.}
\end{center}
\end{figure}

The disorder of a glass-forming liquid implies that flow-event energy barriers $\Delta E$ vary in space throughout the system at any given time. An \textit{ad hoc} assumption is that at any given time the barriers vary randomly in space according to some distribution. This is illustrated in the upper part of \fig{fig:alfabeta}(a) for the case of a distribution, $p(\Delta E)$, which is much wider than $k_BT$. The smallest barriers give rise to what Johari and Goldstein half a century ago termed ``islands of mobility'' \cite{joh70} where fast, spatially isolated rearrangements take place. Such islands may involve just a few molecules or be larger, and they do not necessarily have a super well-defined contrast to the surroundings. On a longer time scale flow events involving larger barriers are gradually ``activated'', and at some point they will percolate the structure. On that time scale extended motion becomes possible within the rigid structure formed by the remaining system. In three dimensions the percolation threshold is roughly one quarter; in two dimensions the threshold is 50\% because a given set or its complement must percolate -- and for geometric reasons both cannot happen \cite{stauffer_percolation,isi92}.

Consider next the largest barriers. Being also spatially isolated, these form ``islands of immobility''. Including gradually smaller barriers, at some point there is slow-domain percolation. This defines a characteristic time scale that we identify with $\ta$: on time scales shorter than $\ta$ the system is rigid and can sustain an externally imposed shear stress, while on longer time scales the slow-domain percolation cluster ``dissolves'', allowing the system to flow in response to an external stress \cite{ton20}. What happens then is that the system flows enough to loosen the structure around any high-energy-barrier location of the liquid. This lowers these barriers before they are transcended, which explains why loss peaks generally are Debye on the low-frequency side. This picture of the $\alpha$ relaxation may be referred to as ``dynamic rigidity percolation'' \cite{dou22}. Note that the long-time flow does not mean the system is like a standard liquid at long times, it is still better regarded as a \stf.

A consequence of the above physical picture is that single-particle motion is predicted to be spatially heterogeneous on short time scales, but homogeneous on time scales longer than $\ta$ \cite{cha97,yam98,ric99,glo00,xu23}. That structural relaxation and thereby $\ta$ is controlled by the slow particles is an old idea \cite{sti88,glo00,kum06,sza06}, which has recently been confirmed in experiments \cite{hig18a}, as well as in equilibrium \cite{gui22,sca22} and aging \cite{dou22} simulations.

The double-percolation picture translates as follows into a generic frequency-dependent loss, $\chi''(\omega)$. The largest barriers are as mentioned never overcome so the corresponding long relaxation times do not contribute to the loss. Consequently, $p(\tau)$ has a long-time cutoff roughly at $\ta$ and $\chi''(\omega)$ is Debye-like at low frequencies: $\chi''(\omega)\propto\omega$ whenever $\omega\ta\ll 1$ \cite{sca22}. On shorter time scales, i.e., above the $\alpha$ loss-peak frequency $\sim 1/\ta$, solidity comes into effect, resulting in an asymmetric loss peak \cite{gui22}. Here we predict $\chi''(\omega)\propto\omega^{-1/2}$ based on solving a simple field theory for the density fluctuations in the Gaussian approximation, assuming a wavevector-dependent density decay rate of the form  $\Gamma(k)=\Gamma_0+\Dc k^2$ in which $\Gamma_0\sim 1/\ta \ll \Dc/ a^2$ \cite{dyr06a,dyr07a}. Conservation laws generally imply a $\Gamma(k)\propto k^2$ dispersion relation arising from the spatial Fourier transform of the $\nabla^2$ operator of the diffusion equation  \cite{kad63,han13}, while the $\Gamma_0$ term corresponds to the apparent violation of density conservation discussed above \cite{dyr06a}. Note that  \fig{fig:alfabeta}(a) relates to the case of a very wide $p(\Delta E)$, which for many equilibrium ultraviscous liquids may apply only at so low temperatures that the system cannot be equilibrated even in long-time experiments. 

A second loss peak is expected at the frequency corresponding to fast-domain percolation (\fig{fig:alfabeta}(a)). Following Gao \textit{et al.} \cite{gao23} we identify this with the ubiquitous Johari-Goldstein (JG) $\beta$ process \cite{joh70}, thereby taking several previous works to their logical consequence \cite{don98,rus00,lon01,ste10,sta13,cic14,yu17,bet18,cap19,cap21,cha22,spi22}. 

A double-percolation picture of glass-forming liquids was proposed already in 1996 by Novikov \textit{et al.}, who discussed percolation of liquid-like and solid-like domains defined by the largest and the smallest vibrational mean-square displacement, respectively \cite{nov96}. A graphic description refers to the slow-domain percolation cluster as a ``sponge'' through which fast motion is possible \footnote{Walter Kob, personal communication (2023)}. 

While we have here focused on linear-response properties of equilibrium glass-forming liquids, it has been demonstrated in colloidal-glass experiments that percolation also controls glass plastic flow; thus growing clusters of nonaffine deformation percolate at yielding \cite{gho17}.

The experimental situation is much less clear than the schematic picture of \fig{fig:alfabeta}(a). In fact, there are only few data for JG $\beta$ relaxation in the equilibrium liquid phase. This is because above $T_g$ the $\alpha$ and $\beta$ processes usually interfere, often to the extent that the high-frequency $\alpha$ decay hides the $\beta$ process that is reduced to an excess wing of the $\alpha$ process \cite{ols98a,sch00b,gui22}. 

Turning to the $\alpha$ process, an analysis of dielectric spectra of 53 liquids revealed that the $\alpha$ high-frequency approximate exponent -- identified as the minimum slope in a log-log plot of the loss, $\am$ -- is predominantly close to -0.5 (\fig{fig:alfabeta}(b)); $\am$ moreover approaches -0.5 as $T\to T_g$ \cite{nie09}. Recent light-scattering data confirm this picture \cite{pab21,sid23}, compare \fig{fig:alfabeta}(c). We also note that recent extensive computer simulations find the exponent -0.38 \cite{sca22}, which is not far from -0.5.

\noindent\textbf{Elastic Models for the Temperature Dependence of $\ta$.}
Point defects in simple crystals are either vacancies or interstitials \cite{ash76,kit76}. Such defects can jump, and the activation energy for a jump scales with the crystal's elastic constants \cite{fly68}. In the \stf picture it is obvious to assume that the flow-event activation energy likewise is proportional to the elastic constants, here those that characterize fast deformations, i.e., the high-frequency plateau shear and bulk moduli. This idea defines the elastic models that exist in several versions \cite{dyr06,rou11}, and which have been linked to models based on the decrease of free volume or increase of collective motion upon cooling \cite{bet15,xu23}. Note that in some models elasticity accounts for only part of the activation energy \cite{mir14a,pha18,mei21}.

\begin{figure}[H]\begin{center}
        \includegraphics[width=4 cm]{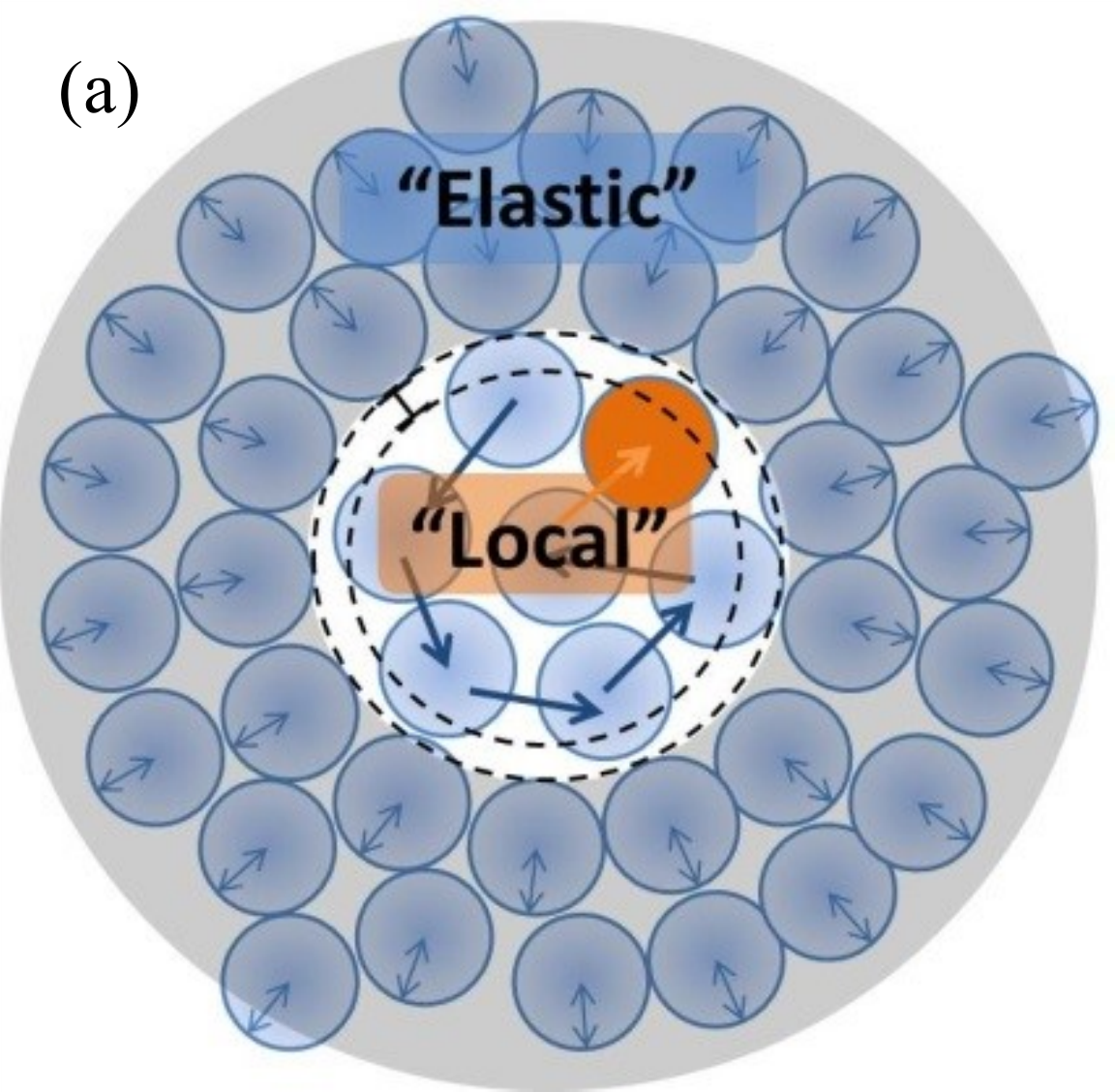}
        \includegraphics[width=5 cm]{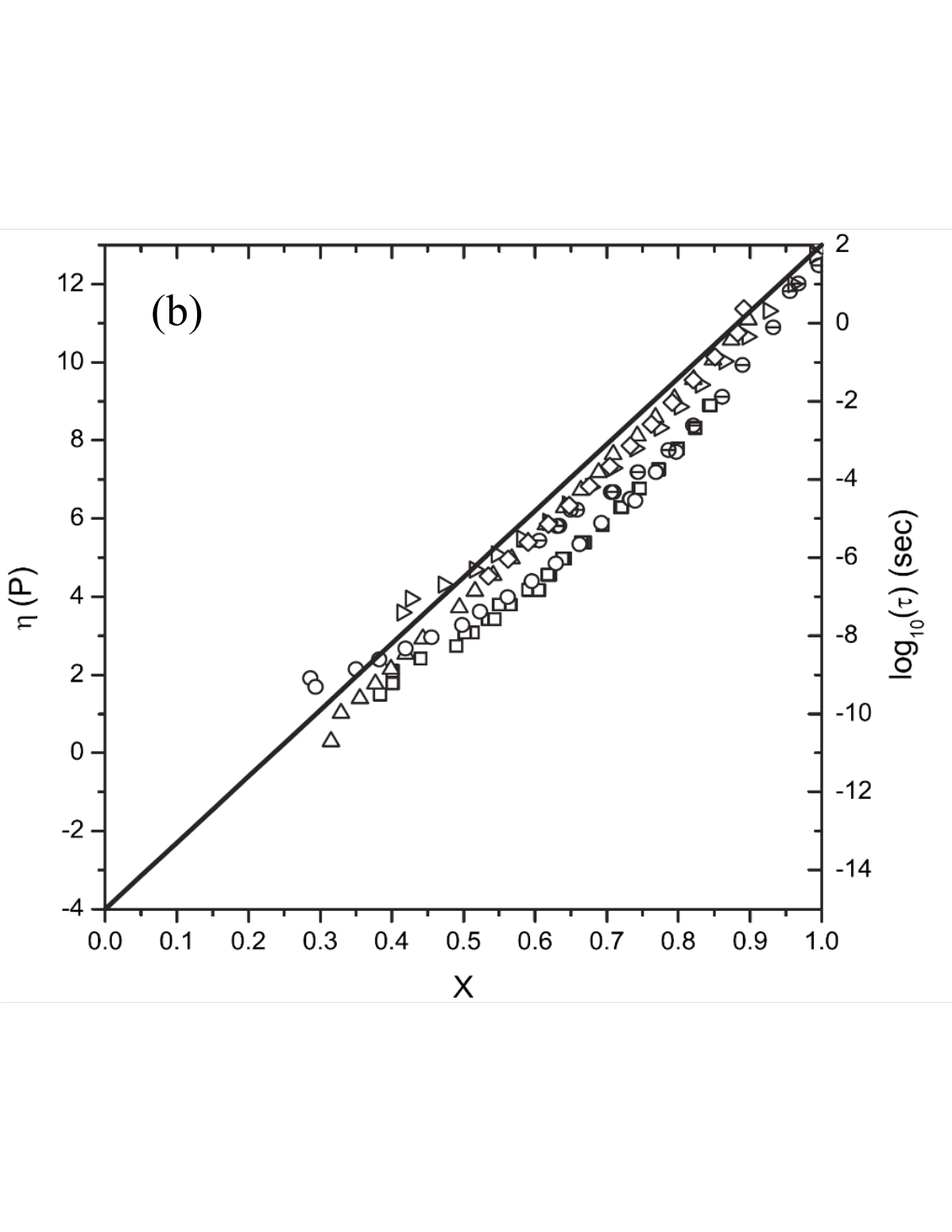}
        \includegraphics[width=5 cm]{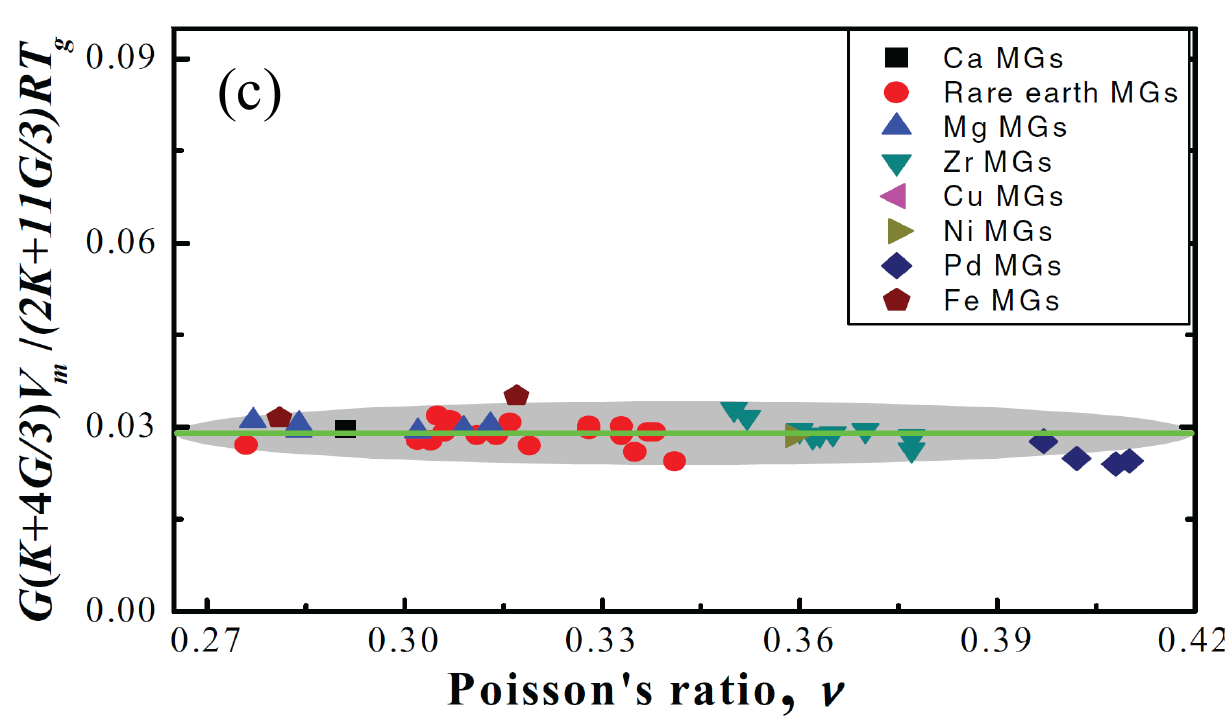}
        \includegraphics[width=4 cm]{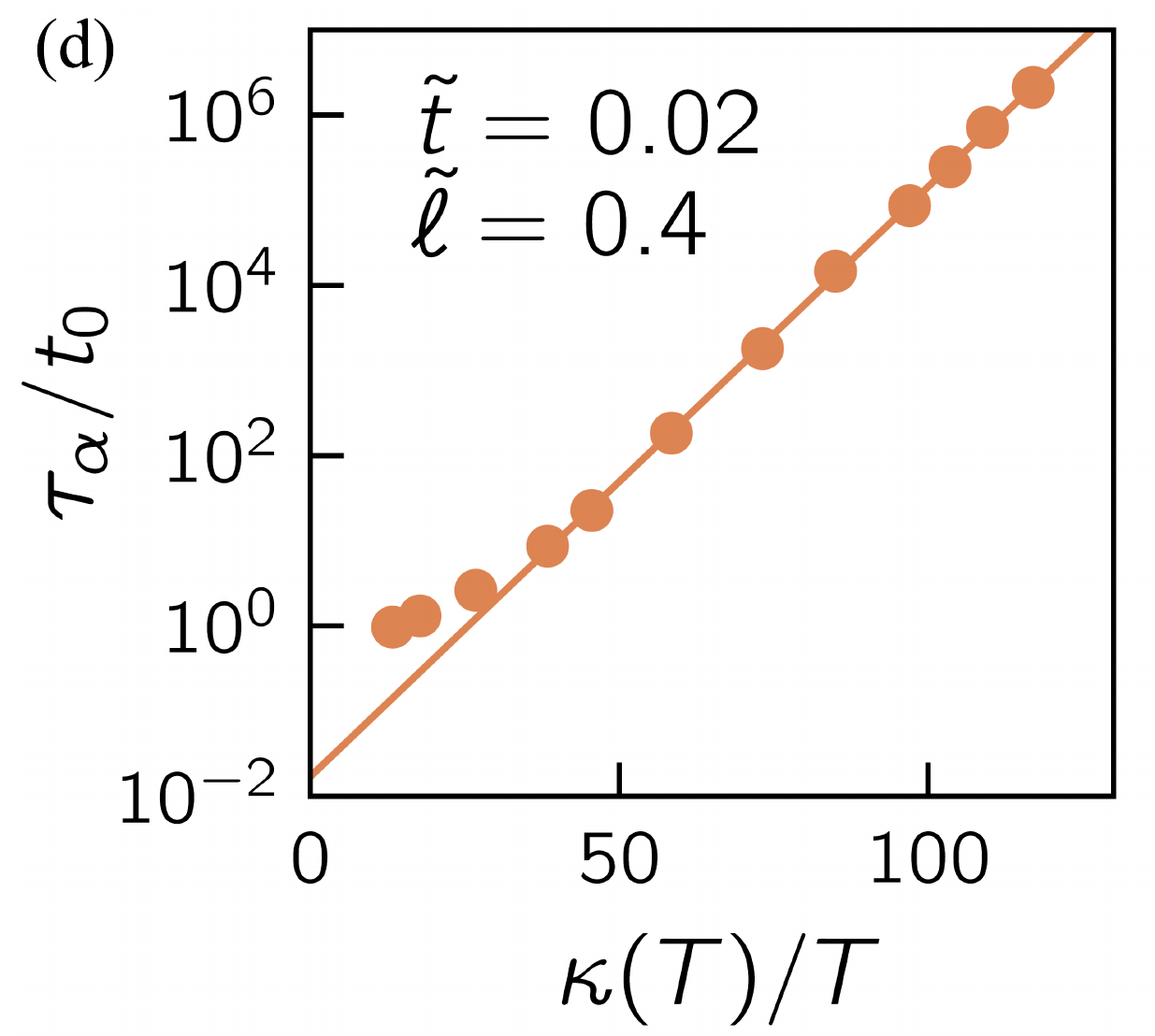}
\end{center}
        \caption{\label{fig:shoving}Elastic model predictions for the temperature dependence of $\ta$.   
        (a) Schematic picture of a flow event. The shoving model for $\ta(T)$ ignores the ``local'' contribution to the activation energy, which is small whenever interactions are strongly anharmonic \cite{dyr98}. 
        Reproduced from Ref. \onlinecite{mir13}. Copyright 2013 American Chemical Society.
        (b) Logarithm of the viscosity of ten organic glass-forming liquids plotted as a function of $X\equiv G_\infty(T)T_g/(G_\infty(T_g)T)$. \Eq{eq:shoving} predicts a straight line ending at the limiting high-temperature viscosity ($10^{-4}$ P =  $10^{-5}$ Pa$\cdot$s). 
        Reproduced from Ref. \onlinecite{tor09}. Copyright 2009 AIP Publishing.
        (c) Test of \eq{eq:mg} for several metallic glasses in which $G$ and $K$ are the shear and bulk moduli of the glass and $V_m$ is the molar volume. 
        Reproduced from Ref. \onlinecite{dyr12}. Copyright 2012 Authors, licensed under a Creative Commons License.
        (d) For a binary Lennard-Jones model the activation energy of $\ta$ is proportional to the average microscopic dipole stiffness $\kappa$, i.e., $\log(\ta/t_0)\propto\kappa(T)/T$ in which $t_0$ is phonon time. 
        Reproduced from Ref. \onlinecite{kap21}. Copyright 2021 AIP Publishing.}
\end{figure}

The simplest mean-field approach assumes that all flow-event activation energies scale in proportion to the macroscopic moduli. For a perfectly spherical flow event in a homogeneous solid, the surroundings experience as mentioned a radial displacement $\propto 1/r^2$. This results in a pure shear deformation, i.e., one with no density change except at the flow event center. Thus the relevant elastic constant is the high-frequency plateau shear modulus $G_\infty$. The shoving model ignores the flow-event center contribution to the activation energy and predicts \cite{dyr96}

\be\label{eq:shoving}
\ta\,=\,\tau_0\,e^{G_\infty(T)V_c/k_BT}\,.
\ee
Here $\tau_0\sim 10^{-13}$s is a prefactor set by the phonon time scale and $V_c$ is a microscopic volume. $G_\infty$ of a glass-forming liquid is usually much more temperature dependent than in the corresponding crystal; in fact $G_\infty(T)$ often increases enough upon cooling to account for the non-Arrhenius $\ta(T)$. The physical picture of the shoving model is given in \fig{fig:shoving}(a). Although many data conform to \eq{eq:shoving} \cite{mei21}, compare \fig{fig:shoving}(b), the shoving model does not apply for all glass-forming liquids  \cite{hec15a,mei21}.

In so far as the dominant contribution to the activation energy derives from displacements around the flow event, not at its center, $G_\infty$ controls more than 90\% of the activation energy \cite{dyr07b}. Elastic models emphasizing instead the bulk modulus also exist, however \cite{var86,mei21}. A popular elastic model expression is $\log(\ta)\propto 1/\langle u^2\rangle$ in which $\langle u^2\rangle$ is the vibrational mean-square displacement \cite{hal87,dyr06,lar08,din16,mei21,lov23}. In this approach $T_g$ is characterized by a definite value of $1/\langle u^2\rangle$, giving rise to a glass analog of the famous Lindemann melting criterion \cite{dud05,ped16,lun23}. This prediction has been investigated by assuming $V_c$ is a system-independent fraction of the molar volume $V_m$ and that all vibrations are phonons controlled by $G_\infty$ and the high-frequency plateau bulk modulus, $K_\infty$. At the glass transition these moduli freeze into their glass values, $G$ and $K$. A straightforward calculation based on the existence of two transverse and one longitudinal phonon for each wavevector leads \cite{dyr04,wan11a,dyr12} to

\be\label{eq:mg}
T_g\propto G\,V_m\frac{K+4G/3}{2K+11G/3}
\ee
with a universal constant of proportionality. This is validated for several bulk metallic glasses in \fig{fig:shoving}(c), thus making it possible to predict $T_g$ from glass properties \cite{sco03}.

The above approaches either assume that the elastic properties are constant throughout the sample or that the local elastic constants \cite{vel18,sha19} scale proportionally when temperature is changed \cite{wei21}. Kapteijns \textit{et al.} studied the energy landscape of a binary Lennard-Jones model to investigate the influence of the elastic constants between neighboring particles on the temperature dependence of $\ta$ \cite{kap21}. The activation energy of $\ta$ was found to be proportional to the average ``stiffness'' between neighboring particles of the liquid’s inherent structures. This leads to the straight-line prediction of \fig{fig:shoving}(d) and suggests an alternative microscopic explanation of why elastic models account for many non-Arrhenius data \cite{hec15a}.

\section*{Discussion}

This perspective has reviewed arguments that a glass-forming liquid below the solidity length is more like a \stf than like ordinary liquids. The focus has been on the dynamics, leaving out a discussion of thermodynamic properties and their correlation to the dynamics \cite{ada65,ang00}. In regard to experimental predictions, we note that the double-percolation picture does not apply in 2d. That is, if $\alpha$ and $\beta$ relaxations as suggested derive from slow- and fast-domain percolation, respectively, no separate Johari-Goldstein $\beta$ processes should exist in 2d because the percolation threshold is here 50\%. It has been argued from simulations that the glass transition in 2d indeed is different from in 3d in several respects \cite{san00,fle15a,ber19}. Another prediction is that all molecules contribute to the $\beta$ relaxation in the liquid phase, albeit only a fraction of them at any given time, while in the glass some molecules contribute and some do not \cite{sca22,vil23}. 

What is the difference between the \stf picture and the standard Maxwell model in which a liquid behaves as a solid over short times? One difference is the existence of a solidity length in the \stf view, a consequence of the fact that the sound velocity is finite, quantifying the length scale above which the picture breaks down. Another difference is that, whereas the Maxwell model predicts standard liquid behavior at long times, the solid-like behavior below the solidity length persists. One consequence of this is the apparent density non-conservation in coarse-grained descriptions.

Not everything discussed in this paper can be correct for the simple reason that there are several inconsistencies. We end the paper by listing these and other issues in order to illustrate that there is still no self-contained picture of glass-forming liquids' solidity and its consequences:

\begin{itemize}

\item The derivation of the solidity length \eq{eq:ls} assumes that each place in the liquid on average gives rise to one flow event every $\ta$. This is inconsistent with the double-percolation picture of \fig{fig:alfabeta}(a) in which a wide range of activation energies is involved. This inconsistency persists even after taking into account that the islands of immobility are  ``renormalized'' by facilitation or otherwise and lowered to the slow-domain percolation activation energy defining $\ta$. This dilemma may be resolved by following Furukawa and instead define $\ls$ as the length scale beyond which ordinary hydrodynamics applies \cite{fur13}; as noted above this leads to virtually the same expression as \eq{eq:ls}. While $\ls\propto\ta^{1/4}$ has recently been confirmed in connection with nonlinear flow modeling \cite{kon22}, it should be mentioned that other recent works predict $\ls\propto\ta^{1/2}$ \cite{vog19} and $\ls\propto\ta^{1/3}$ \cite{fur21}, a matter that needs to be resolved. How to determine $\ls$ in experiments is another important challenge for future work \cite{kon22}.

\item It is not obvious that elastic facilitation (\fig{fig:facilitation}(b) and (c)) is enough to eradicate the largest quarter of the  barriers (\fig{fig:alfabeta}(a)). After all, the stress changes induced by one flow event decay in space as $\propto 1/r^{3}$, which is rather rapid, so other facilitation mechanisms may be needed \cite{sca22,her23}. We favor the above-mentioned possibility that the entire system flows slightly on time scales longer than $\ta$ implying that all stresses decorrelate, including those that keep in place the solid structure defining the local energy barriers.

\item Based on the double-percolation picture one would expect an experimental signature of the percolation critical exponents \cite{gui22,kar23}. This contradicts the prediction that $\chi''(\omega)\propto\omega^{-1/2}$ above the $\alpha$ loss peak frequency. Moreover, it was recently shown that the (zero-parameter) random-barrier model \cite{sch08} provides an excellent fit to the inherent mean-square displacement as a function of time for a binary deeply supercooled Lennard-Jones liquid \cite{sch20}. It is difficult to reconcile that with the prediction for the $\alpha$ high-frequency loss, $\chi''(\omega)\propto\omega^{-1/2}$ \cite{bie17}. 

\item The shoving model assumes uniform elasticity on the short time scale, i.e., does not take into account dynamic heterogeneities. If all flow-event barriers scale proportionally when temperature is changed, this temperature scaling is inherited by $G_\infty(T)$ \cite{wei21}. Even under this assumption, however, one would not expect the flow-event sequences to be temperature invariant because sequences avoiding large barriers become increasingly important as the temperature is lowered. This should lead to a modification of \eq{eq:shoving}.

\end{itemize}

Clearly, much further work is needed before the \stf picture has matured into a simple and coherent one.

\begin{acknowledgments}
The author gratefully thanks the following colleagues for suggestions improving the presentation: Akira Furukawa, Anael Lemaitre, Giulio Biroli, Jack Douglas, Jesper Hansen, Ken Schweizer, Lorenzo Costigliola, Mark Ediger, Massimo Pica Ciamarra, Matthias Fuchs, Misaki Ozawa, Nick Bailey, Peter Harrowell, Roland B{\"o}hmer, Thomas Schr{\o}der, and Walter Kob. Solvej Knudsen and Heine Larsen are thanked for technical assistance.
This work was supported by the VILLUM Foundation's \textit{Matter} grant VIL16515.
\end{acknowledgments}

\end{document}